\documentclass[]{aa} 
\usepackage{graphicx}
\usepackage{txfonts}
\usepackage{url}
\usepackage{natbib}
\usepackage[breaklinks=true]{hyperref} 
\hypersetup{colorlinks=true,urlcolor=blue,citecolor=blue,pdfborder= 0 0 0}
\bibpunct{(}{)}{;}{a}{}{,}    

\begin{document}

 \title{Friends-of-friends galaxy group finder with membership refinement}

 \subtitle{Application to the local Universe}

 \author{E.~Tempel\inst{1}
 	  \and R.~Kipper\inst{1,2}
		 \and A.~Tamm\inst{1}
		 \and M.~Gramann\inst{1}
		 \and M.~Einasto\inst{1}
		 \and T.~Sepp\inst{1,2}
		 \and T.~Tuvikene\inst{1}
   }

 \institute{Tartu Observatory, Observatooriumi~1, 61602 T\~oravere, Estonia\\
    \email{elmo.tempel@to.ee}
   \and
    Institute of Physics, University of Tartu, Ravila 14c, 51010 Tartu, Estonia
    }

 \date{}

 
 \abstract
 {Groups form the most abundant class of galaxy systems. They act as the principal drivers of galaxy evolution and can be used as tracers of the large-scale structure and the underlying cosmology. However, the detection of galaxy groups from galaxy redshift survey data is hampered by several observational limitations.} 
 {We improve the widely used friends-of-friends (FoF) group finding algorithm with membership refinement procedures and apply the method to a combined dataset of galaxies in the local Universe. A major aim of the refinement is to detect subgroups within the FoF groups, enabling a more reliable suppression of the fingers-of-God effect.}
 {The FoF algorithm is often suspected of leaving subsystems of groups and clusters undetected. We used a galaxy sample built of the 2MRS, CF2, and 2M++ survey data comprising nearly 80\,000 galaxies within the local volume of 430~Mpc radius to detect FoF groups. We conducted a multimodality check on the detected groups in search for subgroups. We furthermore refined group membership using the group virial radius and escape velocity to expose unbound galaxies. We used the virial theorem to estimate group masses.}
 {The analysis results in a catalogue of 6282 galaxy groups in the 2MRS sample with two or more members, together with their mass estimates. About half of the initial FoF groups with ten or more members were split into smaller systems with the multimodality check. An interesting comparison to our detected groups is provided by another group catalogue that is based on similar data but a completely different methodology. Two thirds of the groups are identical or very similar. Differences mostly concern the smallest and largest of these other groups, the former sometimes missing and the latter being divided into subsystems in our catalogue.}
 {}


 \maketitle
%
\section{Introduction}

Galaxies may reside in an extraordinary variety of environments of different scale and nature. Moreover, a galaxy can be embedded in different environment types at the same time -- consider a pair of galaxies, each with its satellite system, inhabiting a large-scale filament within a supercluster.

The exact definition for a galaxy group or a cluster (we use the term `group' for both throughout) tends to vary from author to author and, even worse, from system to system. Nevertheless, we can take it for granted that the group is the primary level of environment for any given galaxy and has the most direct effect on the evolution of the galaxy. It also is the main receiver of feedback from the galaxy's gravitational potential, radiation, galactic winds, AGN jets, etc. This mutual relationship suggests that galaxy group catalogues provide an indispensable tool for studying galaxy evolution. On the other hand, galaxy groups and clusters are the largest gravitationally bound systems (by the typical definition) and are tracers and characterisers of the cosmic web of voids, sheets, filaments, and superclusters. Therefore, galaxy group catalogues essentially provide handy means for estimating various cosmological parameters, their evolution and interrelation, and for validating cosmological simulations.

The problem is that no straightforward procedure exists for determining a galaxy group from observational data. Out of the six real- and velocity-space coordinates required to decide whether a galaxy does or does not belong to a given group, only three are provided by redshift surveys. Most importantly, redshift measurements cannot distinguish peculiar motions of galaxies from their drift along the Hubble flow. The resulting stretching of galaxy groups in the redshift space, the fingers-of-God effect\footnote{The term coined at the IAU Symposium~79 in 1977 in Tallinn by \citet{1978IAUS...79...31T}; the effect itself was first noted by \citet{1972MNRAS.156P...1J}.}, makes it difficult to mark group boundaries in the radial direction. 

Over the years, the community has developed an arsenal of algorithms to overcome the incompleteness of data for group detection, which, in one way or another, make assumptions about the gravitational potential and the 3D shape of groups. \citet{2014MNRAS.441.1513O, 2015MNRAS.449.1897O} compared the performance of many of these methods on mock observational data. A trivial but nonetheless important conclusion was made: the recovery of group properties most critically depends on the accuracy of membership determination.

From the scientific point of view, registries of galaxy systems in the local Universe are of particular importance. In our cosmological neighbourhood, fainter and smaller galaxies are visible, more and higher quality data are available for any further analysis, redshift-independent distance estimators are available for many sources, etc. The Two Micron All Sky Survey and its extensions \citep[2MASS;][]{2000AJ....119.2498J,2006AJ....131.1163S,2011MNRAS.416.2840L,2012ApJS..199...26H} offer a so far unrivalled dataset of galaxies in the local Universe that allows cataloguing galaxies across most of the celestial sphere. Using these data, \citet{2003ApJ...585..161K} compiled a catalogue of galaxy clusters using the `matched-filter' algorithm, in which clusters are identified as overdensities with respect to a background distribution. \citet{2007ApJ...658..917D} applied the same algorithm to compile a catalogue of galaxy clusters and study their X-ray properties and baryon fractions \citep{2010ApJ...719..119D}, while \citet{2007ApJ...655..790C} and \citet{2011MNRAS.416.2840L} applied variations of the popular `friends-of-friends' (FoF) algorithm to detect galaxy groups. \citet{2012MNRAS.426..296D} presented a photometric catalogue of compact groups of 2MASS galaxies and studied their properties, especially the properties of their first-ranked galaxies. Most recently, \citet{2016MNRAS.455.3169L} used a Bayesian approach to extract structures from the galaxy distribution, while \citet{ 2015AJ....149...54T} introduced the power of scaling relations to constrain galaxy groups and applied it on 2MASS galaxies to construct a group catalogue \citep{2015AJ....149..171T}.

In this paper, we present a catalogue of galaxy groups in the nearby Universe. The groups and clusters have been recovered by applying the FoF method, improved according to the lessons learned from studies of the substructure of groups with the \emph{mclust} package \citep{2010A&A...522A..92E, 2012A&A...540A.123E, 2013MNRAS.434..784R}. Our product is mainly based on the 2MASS Redshift Survey. This enables a straight comparison with the latest catalogues that rely on the same data, but are fundamentally different group detection principles. In addition to cross-checking the group finder algorithms, this comparison also allows us to characterise the 2MASS data as a basis for galaxy group studies.

Throughout this paper we assume the Planck cosmology \citep{2015arXiv150201589P}: the Hubble constant $H_0 = 67.8~\mathrm{km~s^{-1}Mpc^{-1}}$, the matter density $\Omega_\mathrm{m} = 0.308$, and the dark energy density $\Omega_\Lambda = 0.692$.

\section{Galaxy data}

\begin{figure}
 \centering
 \includegraphics[width=88mm]{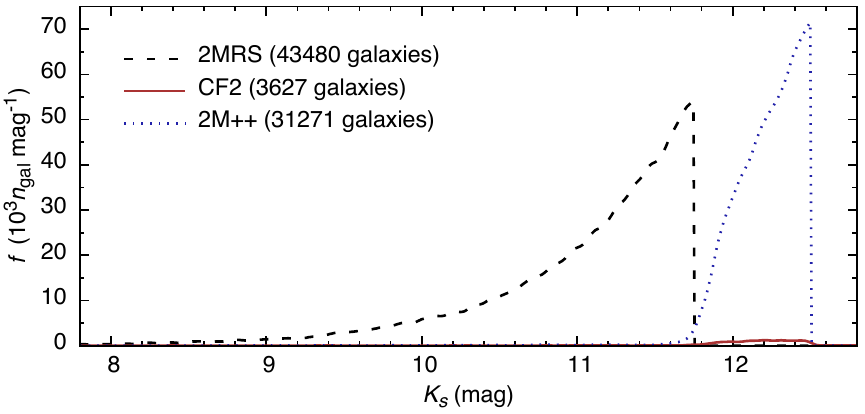}
 \caption{Observed magnitude distribution for 2MRS, CF2, and 2M++ datasets in our final galaxy sample. Here, only those CF2 and 2M++ galaxies are considered that are not present in the 2MRS catalogue. The corresponding numbers are shown in the legend.}
 \label{fig:magnitude_distribution}
\end{figure}

To delineate galaxy groups in the local Universe, we used galaxy data from the extragalactic distance database \citep[EDD\footnote{\url{http://edd.ifa.hawaii.edu}.};][]{2009AJ....138..323T}. The sample encompasses three datasets. As the main source, we used the Two Micron All Sky Survey \citep[][]{2006AJ....131.1163S} Redshift Survey (2MRS) galaxies brighter than 11.75 magnitudes in the $K_s$ band \citep[for a description of the catalogue, see][]{2012ApJS..199...26H}. We only used galaxies that are securely off the Galactic plane: Galactic latitude $|b|>5\degr$. Since the galaxy sample becomes extremely sparse farther away, we only used galaxies with a cosmic microwave background (CMB) corrected redshift $z=0 \dots 0.1$ (up to 430~Mpc). This selection restricts our 2MRS sample to 43480 galaxies.

For our analysis, we complemented the main 2MRS sample with two other sources. From the CosmicFlows-2 survey that contains 8198 galaxies with redshift-independent distance estimates \citep[CF2;][]{2013AJ....146...86T}, we added 3627 (of these, 2799 galaxies do not have a measured $K_s$ magnitude). In addition, we made use of the 2M++ catalogue \citet{2011MNRAS.416.2840L}, which combines elements from the 2MRS, the 6DF Galaxy Survey \citep[][]{2009MNRAS.399..683J}, and the Sloan Digital Sky Survey \citep[][]{2000AJ....120.1579Y}. Of the 64745 galaxies of the 2M++, we added 31271 galaxies\footnote{The actual number at the time of our download; the original paper gives a different number.} down to $K_s<12.5$ \footnote{We note that the 2M++ magnitudes are defined slightly differently from the 2MRS magnitudes \citep[see][for details]{2011MNRAS.416.2840L}. Fortunately, our group construction algorithm does not depend on galaxy luminosities.}, which extends the sample well beyond the 2MRS magnitude limit.

Our final galaxy dataset includes 78378 galaxies. The observed magnitude distribution for the 2MRS, CF2, and 2M++ subsamples is shown in Fig.~\ref{fig:magnitude_distribution}. The dataset is incomplete for galaxies with $K_s>11.75$. This should be taken into account when strictly flux- or volume-limited samples are needed (e.g. for constructing galaxy luminosity functions or calculating the luminosity density field). The incompleteness is not a serious problem in the catalogue construction because the membership was individually refined for each group (see Sect.~\ref{sec:fof_member_refinement}). Moreover, the dataset is not complete for the nearby Universe, where many dwarf galaxies cross the magnitude threshold, but are missed by the 2MASS survey because of their low surface brightness \citep{2013AJ....145..101K}. The best group catalogue for the nearby ($d \lesssim 40$~Mpc) Universe is probably the one constructed by \citet{2011MNRAS.412.2498M}.

\begin{figure}
 \centering
 \includegraphics[width=88mm]{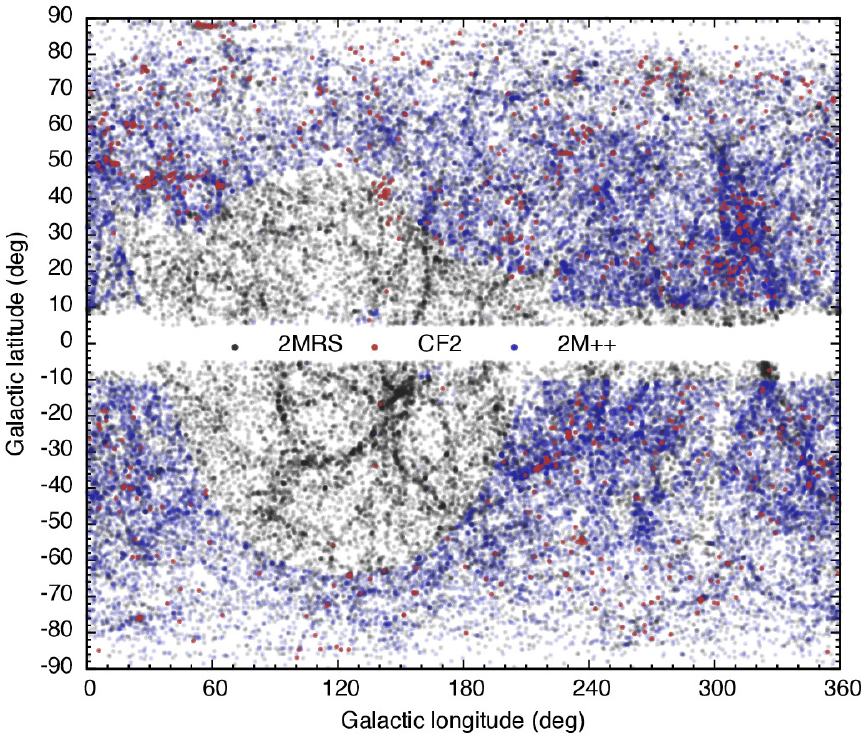}
 \caption{Sky distribution of galaxies in the 2MRS, CF2, and 2M++ datasets. Of the CF2 and 2M++ datasets, only galaxies not present in the 2MRS catalogue are shown. See Fig.~3 in \citet{2011MNRAS.416.2840L} for the sky coverage in the 2M++ dataset.}
 \label{fig:sky_distribution}
\end{figure}

\begin{figure}
 \centering
 \includegraphics[width=88mm]{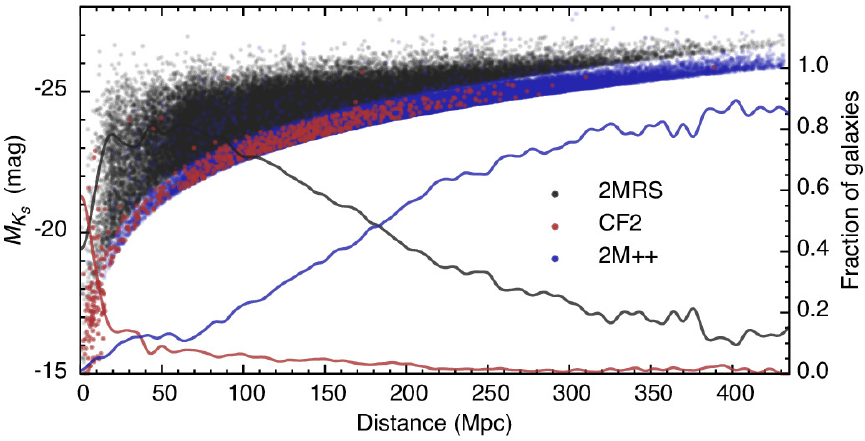}
 \caption{Galaxy absolute magnitude as a function of distance in 2MRS, CF2, and 2M++ datasets. Here, the CF2 and 2M++ datasets include only galaxies that are not present in 2MRS catalogue. Solid lines show the fraction of galaxies in the complete sample as a function of distance.}
 \label{fig:distance_vs_mag}
\end{figure}

Figure~\ref{fig:sky_distribution} shows the distribution of galaxies in the plane of the sky. While the CF2 and 2MRS galaxies are distributed all over the sky, those of the 2M++ are not. The completeness of the 2M++ sample is fully described in \citet{2011MNRAS.416.2840L}. Figure~\ref{fig:distance_vs_mag} shows the luminosities of galaxies and the relative contributions by the 2MRS, CF2, and 2M++ subsamples as a function of distance. The 2MRS provides the bulk of the galaxies in the nearby region and is supplemented by the CF2, while the 2M++ becomes dominant farther away. These two catalogues combined provide a representative galaxy sample up to 400~Mpc. With a higher number density of galaxies we gain more reliable groups with more reliable properties, which means a better input for any subsequent analysis. For example, we intended to use the prepared dataset to extract galaxy filaments from the local Universe using the Bisous model \citep{2014MNRAS.438.3465T}. For the filament detection, a high number density of galaxies is preferred while the varying completeness in the sky is not a concern.

Considering the above, the group construction and the resulting catalogue are presented separately for two cases: the full dataset and the pure 2MRS, the latter being more suitable for studies where completeness is of critical importance.

\section{Group detection and membership refinement}

\subsection{Conventional friends-of-friends group finder}
\label{sec:conventional}

One of the simplest and therefore most widely used algorithms for group detection is the FoF percolation method\footnote{Also known as single-linkage clustering among statisticians.}, as first reported by \citet{1976ApJS...32..409T}, \citet{1982ApJ...257..423H}, and \citet{1982Natur.300..407Z}. We used the FoF method previously to detect galaxy groups from SDSS redshift-space catalogues \citep{2008A&A...479..927T, 2010A&A...514A.102T, 2012A&A...540A.106T, 2014A&A...566A...1T}.

Recently, \citet{2014MNRAS.441.1513O, 2015MNRAS.449.1897O} inspected in detail the ability of various group detection algorithms to recover the actual groups and group masses using simulated galaxy catalogues. The results indicated that using the standard calibration for the linking length (see below), the FoF method recovers galaxy groups reasonably well.

When the FoF method is applied to redshift-space catalogues of galaxies, the only free parameters are the linking lengths in radial ($b_{||}$, along the line of sight) and in transversal ($b_{\perp}$, in the plane of the sky) directions. These linking lengths are often calibrated according to simulations \citep[see e.g.][]{2004MNRAS.348..866E, 2011MNRAS.416.2640R}; values close to $b_{\perp}\approx 0.1$ and $b_{||}\approx 1.0$ in units of mean separation between galaxies are typically used. A detailed analysis of how linking length values affect the detected galaxy groups has been conducted by \citet{2014MNRAS.440.1763D}. They suggested that $b_{\perp}\approx 0.07$ and $b_{||}\approx 1.1$ (or even higher in radial directions, depending on the goal of the study) should be used.

\citet{2014A&A...566A...1T} proposed that for redshift-based catalogues, the linking length should be calibrated according to the mean distance to the nearest galaxy (i.e. the mean separation) in the plane of the sky. The distance should be measured considering the nearest (in sky projection) neighbour within a cylindrical volume defined by the same fixed $b_{||}/b_{\perp}$ ratio as used in the subsequent FoF analysis.

\begin{figure}
 \centering
 \includegraphics[width=88mm]{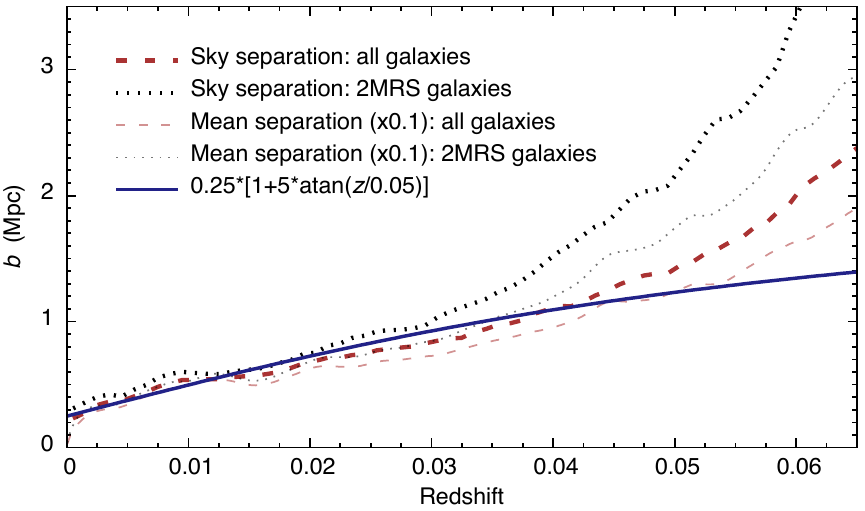}
 \caption{Mean separation between galaxies in the plane of the sky (thick lines) and in 3D space (thin lines) as functions of redshift. To find the sky projection distances, the nearest neighbour was sought within a volume defined by the fixed linking length ratios of the FoF algorithm (see text for details). Dashed lines correspond to the whole sample, dotted lines to the 2MRS galaxies alone. Solid blue line shows the transversal linking length as used in our group finding algorithm. In the nearby region, the transversal linking length is roughly 0.1 times the mean separation of galaxies of the given samples. We note that the galaxy sample reaches redshift $z=0.1$; for illustrative reasons, the figure only presents the nearer part.}
 \label{fig:linking_length}
\end{figure}

Figure~\ref{fig:linking_length} shows that the mean distance to the nearest galaxy in the plane of the sky corresponds well to the actual 3D mean separation between galaxies. However, the actual 3D separation cannot be directly used to calibrate the linking length because of redshift space distortions. To overcome this problem, distances in the radial direction need to be multiplied by the same $b_{||}/b_{\perp}$ ratio as is used for FoF analysis during the calculation of the mean separation. The result would be effectively identical to our current approach. 

For flux-limited surveys, the FoF linking length should increase with distance because fainter galaxies are not detected farther away. \citet{2014A&A...566A...1T} scaled the linking length according to nearby groups and found a correction function to take the dropping out of fainter group members with increasing distance into account. This scaling with distance is well expressed with an arctangent law. For the current sample the dependence of the linking length (in the transversal direction) on $z$ can be expressed as
\begin{equation}
	\frac{b_{\perp}(z)}{\mathrm{Mpc}} = 0.25\left[1+5\arctan\left(z/0.05\right)\right],
	\label{eq:arctan}
\end{equation}
which is also plotted in Fig.~\ref{fig:linking_length}. We used the linking length in physical and not in comoving units. However, the difference is negligible for the given redshift range.

 This scaling correlates very well with the mean separation up to redshift 0.04 (see Fig.~\ref{fig:linking_length}), while farther away the mean separation increases faster than the arctan law. The discrepancy emerges when all the other members of the group remain undetected as a result of the flux limit, and for a given galaxy, the nearest galaxy is found from some neighbouring group.

Compared to \citet{2014MNRAS.440.1763D}, our linking length in transversal directions is slightly higher (0.1 vs 0.07 times the mean separation up to redshift 0.04). Nevertheless, we continued to use our value for the following reasons: it has worked well in our previous catalogues and in the comparison project \citep{2014MNRAS.441.1513O, 2015MNRAS.449.1897O}; we conducted a subsequent membership refinement; and following \citet{2014MNRAS.440.1763D}, we can conclude that at the given level, the effect of linking length differences on the results are marginal.

The remaining question in our FoF method implementation is the choice of the radial linking length $b_{||}$. So far, no clear-cut recipe exists. With too low a value we would miss group members with high peculiar velocities and thus also underestimate group masses. Too high a value would contaminate the detected groups with outliers and merge separate groups. In our previous papers, we have found the balance using $b_{||}/b_{\perp}=10$, while for example \citet{2014MNRAS.440.1763D} proposed that $b_{||}/b_{\perp}\approx 16$ or even higher should be used. In the following, we conservatively raise our previously used value to $b_{||}/b_{\perp}=12$ to gain more group members in the radial direction. A potential contamination would be reduced later with the membership refinement procedure. 

A further complication with the linking length is that in principle, it depends on the underlying environment density \citep[see e.g.][]{2004MNRAS.348..866E,2011MNRAS.416.2640R}. However, because the dependency is weak and can thus only slightly affect the FoF group detection, we did not adjust the linking length according to density, but relied on the membership refinement in this aspect as well.

Figure~\ref{fig:linking_length} shows that the mean separation is roughly the same for 2MRS and the whole dataset up to redshift 0.04. Hence, we can use the same FoF linking length values and scaling in both cases.

\subsection{Friends-of-friends group member refinement}
\label{sec:fof_member_refinement}

The conventional FoF group finder is simple and works reasonably well in most situations, but it also has its drawbacks. If two groups are merging or they simply happen to lie too close to each other, the FoF algorithm may detect them as a single system. Additionally, FoF groups can become ``hairy'', meaning that near the outer edges of groups the surrounding field galaxies are considered as group members. Galaxy filaments connected to groups can also be mistaken for group members by the FoF algorithm.

As pointed out by \citet{2014MNRAS.441.1513O, 2015MNRAS.449.1897O}, even a simple membership refinement after an initial FoF group detection can significantly enhance the reliability of the groups. Here we conducted the refinement in two steps. First, we used a multimodality analysis (see Sect.~\ref{sec:mclust}) to detect multi-component groups and to split them into independent systems. Second, we used estimates of the virial radius and the escape velocity to exclude group members that are not physically bound to groups (see Sect.~\ref{sec:member_refinement}).

\subsubsection{Group refinement using multimodality analysis}
\label{sec:mclust}

\begin{figure}
 \centering
 \includegraphics[width=88mm]{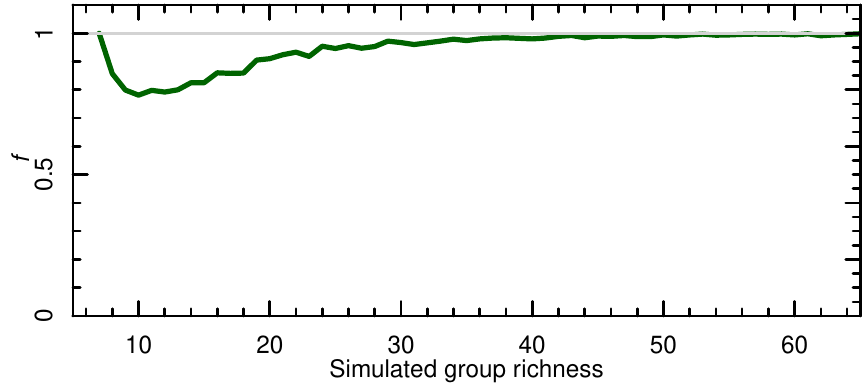}
 \caption{Fraction of correct unimodality detections in intrinsically unimodal Gaussian mock groups as a function of group richness. For groups with 8--20 members, the \emph{mclust} algorithm returns an incorrect multimodality detection in about 10--20\% of the cases.}
 \label{fig:frac_model}
\end{figure}

\begin{figure*}
\sidecaption
\includegraphics[width=12cm]{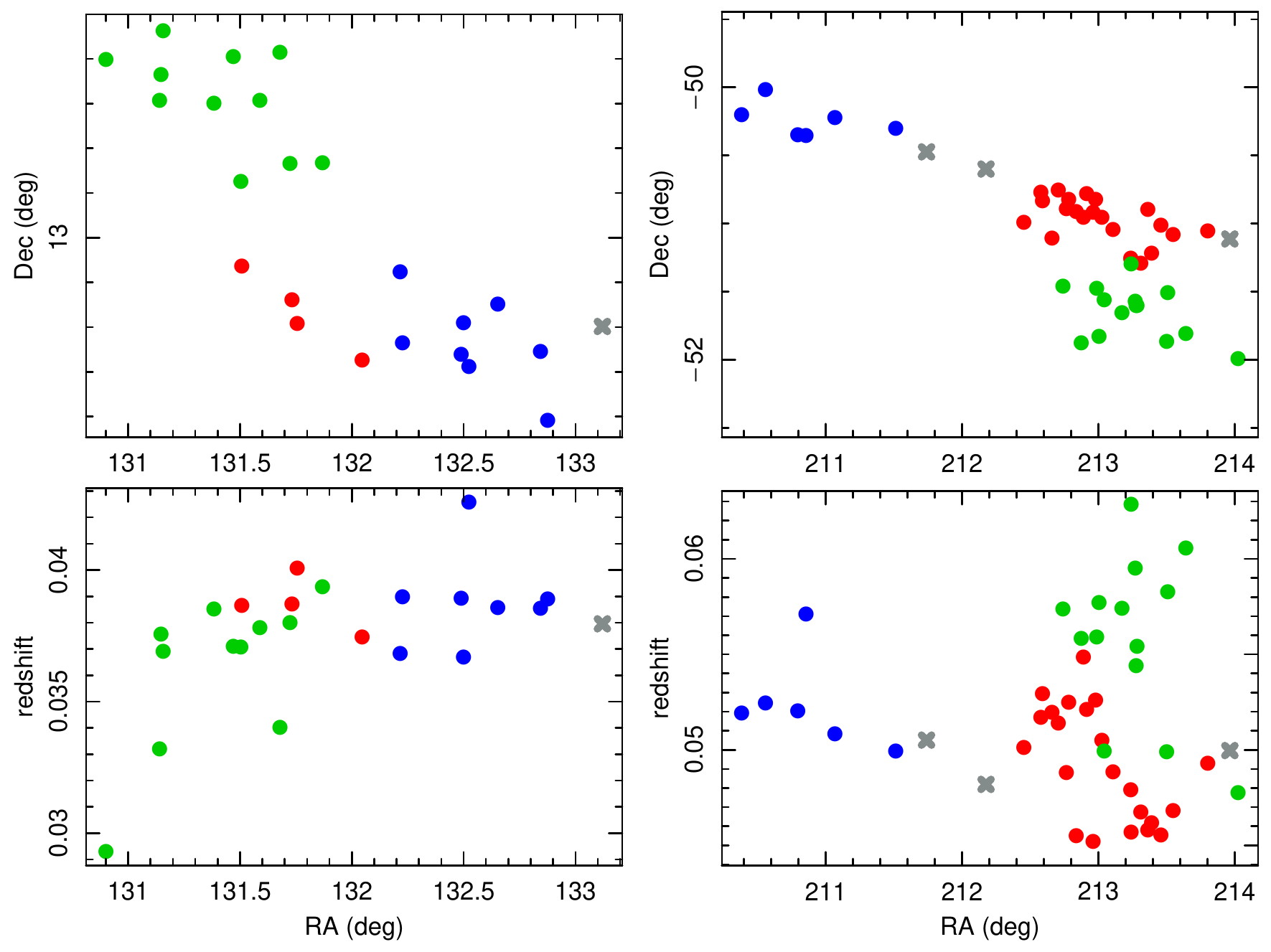}
\caption{Example of two galaxy groups (in \emph{left-} and \emph{right-hand panels}, respectively) as detected by the FoF algorithm. Galaxy positions are shown in RA-Dec (\emph{upper panels}) and redshift-Dec (\emph{lower panels}) coordinates. The conventional FoF method sees both systems as single groups, while the multimodality analysis has split them into subsystems, indicated with different colours (see Sect.~\ref{sec:mclust} for details). Grey crosses represent galaxies that do not belong to any subsystem; these galaxies were removed during the final group membership refinement, as explained in Sect.~\ref{sec:member_refinement}. The \emph{right-hand panels} show that the blue subgroup is connected with the others by the initial FoF detection because of the two central galaxies, which were identified as outliers in the subsequent analysis. The figure illustrates that membership refinement may be of critical importance for richer groups detected with FoF algorithm.}
\label{fig:mclust}
\end{figure*}

To check the multimodality of groups found by the FoF algorithm, we used a model-based clustering analysis assuming a Gaussian distribution for the number density of group members. The method is implemented in the statistical computing environment \emph{R}\footnote{\url{https://www.r-project.org}.} in the package \emph{mclust} \citep{mclust1, mclust2}. For the clustering analysis we fixed the expected number of subgroups and using the EM (expectation maximisation) algorithm, \emph{mclust} finds the most probable locations, sizes, and shapes for each subgroup. Additionally, \emph{mclust} gives the probability of each galaxy of being within each subgroup. In the end, each galaxy is assigned to a single group according to the highest probability. We only applied the \emph{mclust} analysis on systems with at least seven galaxies.

Since in galaxy redshift surveys, groups are not spherical but elongated along the line of sight due to the FoG effect, we fixed one axis with the line of sight during the clustering analysis. The other two axes were set perpendicular with the line of sight and each other, while the orientation in the sky plane was left free.

We ran \emph{mclust} with different expected numbers of subsystems (from one to ten) to determine the most probable value. The latter was then chosen using the Bayesian information criterion (BIC), which is widely used in statistics and is implemented in the \emph{mclust} package.

The clustering algorithm was applied on each FoF group separately. If the algorithm detected subcomponents, we ran the same algorithm on each subcomponent to test whether even more substructure could be found. In most cases, the first run was sufficient; a more detailed analysis only affects large FoF clusters where the instant detection of subsystems is complicated.

The multimodality analysis might detect subgroups as a result of pure spatial coincidence of galaxies, especially in smaller groups. To estimate the level of this uncertainty, we performed the following test. For each group richness, we generated 1000 Gaussian groups whose distribution was elongated along one coordinate axis to mimic the FoG effect. For each simulated group we ran the \emph{mclust} algorithm as we did for the observed sample and estimated the fraction of false detections of multimodal systems. The results are shown in Fig.~\ref{fig:frac_model}. For systems with about ten member galaxies, the false detection rate is the highest, being around 20\%. For groups with 20 members or more, this fraction falls below 10\%. This estimate agrees with the one made by \citet{2013MNRAS.434..784R}. We therefore conclude that compared to the expected observational selection effects, the additional uncertainty of group membership arising from the multimodality check is small.

\subsubsection{Group membership refinement using virial radius and escape velocity}
\label{sec:member_refinement}

As the final step in our group construction, we removed all galaxies that were apparently not bound to the systems, either according to the virial radius or to the escape velocity of the group. Thus a galaxy was excluded from its group if its distance from the group centre in the plane of the sky was greater than the virial radius. The group centre was calculated as the geometrical centre of all galaxies in the group without any luminosity or mass weighting. As the virial radius we took $r_{200}$, the radius of a sphere in which the mean matter density is 200 times higher than the mean of the Universe. The value of $r_{200}$ is entirely determined by the virial mass, which we estimated using the virial theorem assuming an NFW mass density profile as described in \citet{2014A&A...566A...1T} and in Appendix~\ref{app:properties}. In brief, we used the velocity dispersion and the projected gravitational radius to estimate a group's mass via the virial theorem. The mass estimation is thus fully described by the theory and does not require any scaling parameters. The underlying calculations of group velocity dispersions and sizes in the plane of the sky are described in Appendix~\ref{app:properties}. 

Similarly, a galaxy was removed from its group if the velocity of the galaxy with respect to the group centre was higher than the escape velocity at its sky-projected distance from the group centre.
The escape velocity of a group relates to the gravitational potential $\Phi$ through
\begin{equation}
	v^2_\mathrm{esc}(r) = -2\Phi(r) ,
\end{equation}
where $r$ is distance from the group centre. Gravitational potential is directly related to the assumed dark matter density profile \citep[see e.g.][]{2001MNRAS.321..155L}. We note that our approach is conservative: the sky-projected distance generally underestimates the 3D distance, thus we tend to overestimate the escape velocity, leaving some outliers in the group rather than removing real members.

The member refinement for groups was done iteratively since the group velocity dispersion and size obviously depend on the group membership. For most of the groups, the refinement converged after a few iterations. Since group masses cannot be estimated for small groups, we applied the outlier detection only on groups with at least five member galaxies.

In some cases the excluded members form separate compact systems of a few members at the boundaries of larger groups. To detect these systems as groups, we reran the full group detection and membership refinement procedure on the excluded members. With this iterative approach, we detected small groups that had remained undetected during the multimodality analysis, but were revealed during the membership refinement according to the virial radius and escape velocity.

\begin{figure}
 \centering
 \includegraphics[width=88mm]{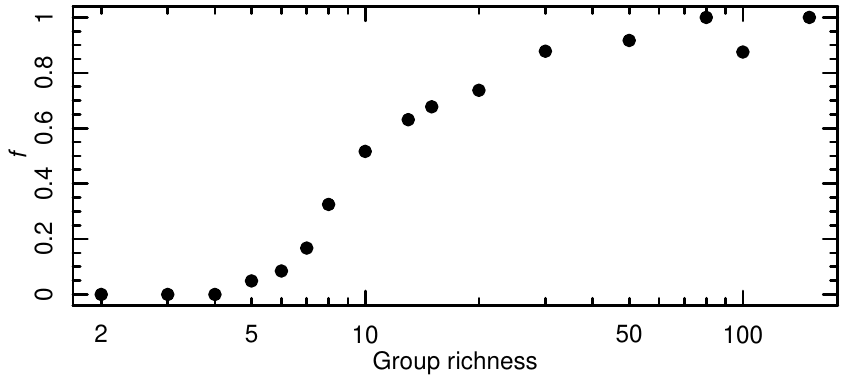}
 \caption{Fraction of groups that were split into multiple components during group membership refinement,  shown as a function of the initial FoF group richness.}
 \label{fig:frac_multimodal}
\end{figure}

\section{Group finder in action}
\label{sec:cat}

Now we applied the galaxy group construction algorithm explained in the previous section to galaxy redshift surveys of the local Universe. We constructed galaxy groups separately for two datasets: the 2MRS and the combined 2MRS, CF2, 2M++ datasets, detecting 6282 and 12106 groups, respectively, with two or more members. Since the combined dataset is roughly twice as large as the 2MRS dataset, the similar difference of the number of the detected groups is expected. The fraction of galaxies in groups in the 2MRS alone and in the combined dataset is 45\% and 50\%, respectively. A description of the corresponding group catalogues is given in Appendix~\ref{app:cat}.

In general, the FoF method is reliable \citep[see e.g.][]{2014MNRAS.441.1513O, 2015MNRAS.449.1897O} and statistically, further refinement affects a relatively small fraction of groups. Of the initial FoF groups, roughly 2\% were detected as multi-component systems in the given datasets. As expected, these were mostly among the largest systems (see Fig.~\ref{fig:frac_multimodal}). On the other hand, about half of the systems with at least ten galaxies were affected -- hence, the multimodality analysis is an important addition to the traditional FoF algorithm for larger systems. Outlier identification using virial radius and escape velocity was necessary for about 10\% of the systems. Once again, systems with more members benefited from the refinement more.

A visual inspection confirms that galaxy systems split into smaller groups by the multimodality analysis contain apparent substructure. Similarly, a random check reveals that the removed outliers appear not to be tightly connected to the groups. Figure~\ref{fig:mclust} shows examples of these cases. From the right-hand panels we can deduce that the FoF algorithm considers the given galaxy system as a single group because of the two galaxies at the centre of the figure. The membership refinement (see also Sect.~\ref{sec:member_refinement}) suggests that these two galaxies do not belong to any of the subgroups. Figure~\ref{fig:mclust} also shows that the multimodality analysis can separate nearby (potentially merging) groups that are clearly distinct in the sky plane and/or in the redshift space.

\begin{figure}
 \centering
 \includegraphics[width=88mm]{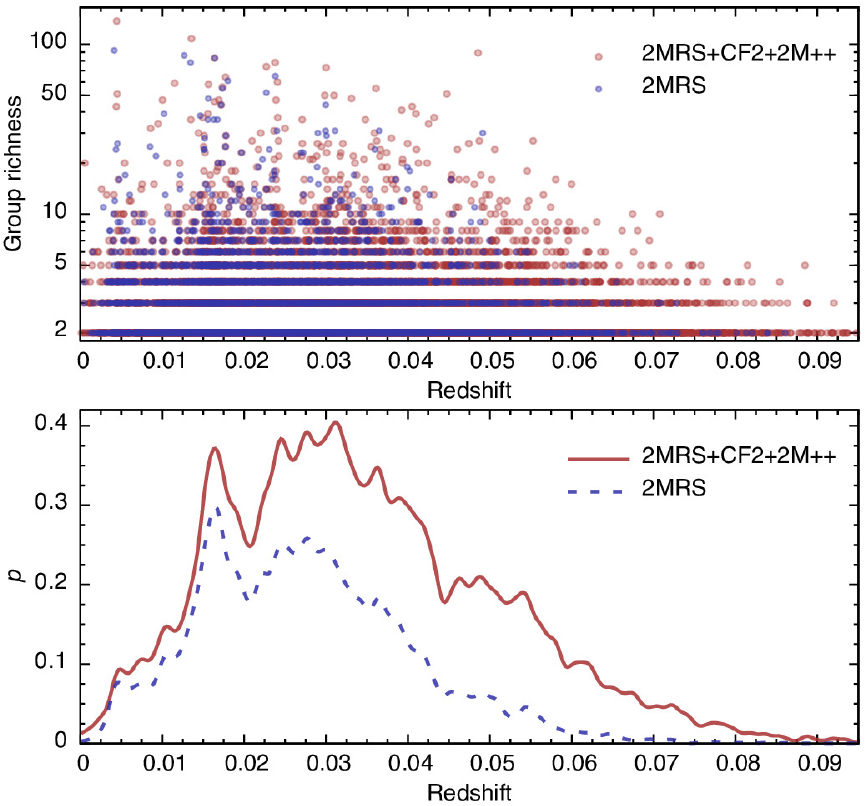}
 \caption{\emph{Upper panel:} galaxy group richness (number of galaxies in a group) as a function of redshift. \emph{Lower panel:} redshift distribution of groups for the 2MRS sample and for the combined dataset. The lack of richer groups farther away is due to the flux limit of the data.}
 \label{fig:dist_vs_ngal}
\end{figure}

\begin{figure}
 \centering
 \includegraphics[width=88mm]{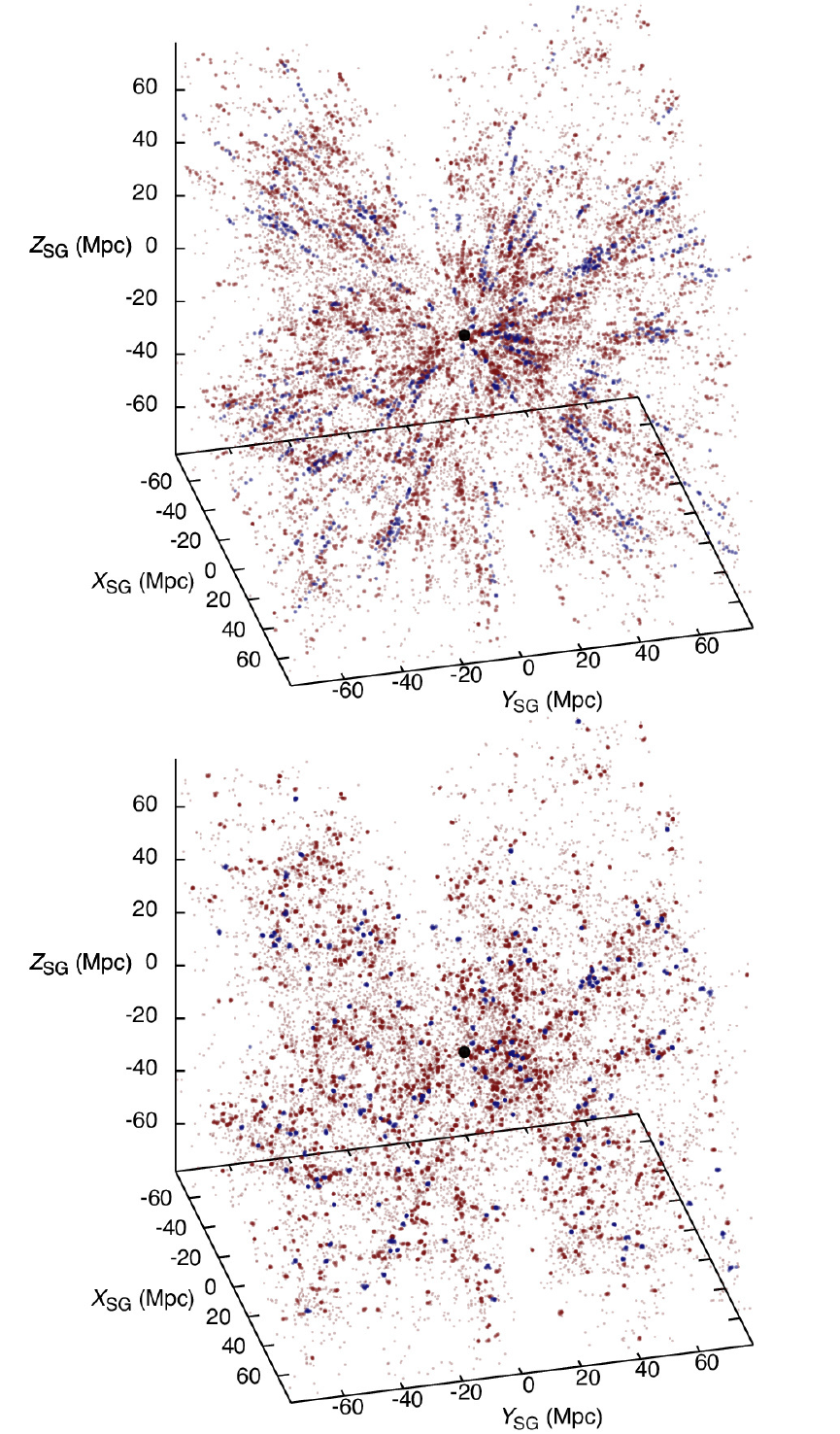}
 \caption{Galaxy distribution in the local Universe according to the combined dataset, presented in comoving supergalactic cartesian coordinates. The observer is located in the centre of the figure, marked with a black point. In the \emph{upper panel}, the observed galaxy distribution with redshift-based distances is shown. Galaxies in groups with more than five members are shown as blue points, other galaxies as red points. To emphasise the FoG effect, isolated galaxies and galaxy pairs are shown with slightly smaller points. In the \emph{upper panel} elongated structures (fingers of God) are clearly visible along the line of sight -- the galaxy distribution points towards the centre of the figure. In the \emph{lower panel} the same galaxies are plotted after the FoG effect was suppressed as described in the text.}
 \label{fig:fog}
\end{figure}

\begin{figure*}
 \sidecaption
 \includegraphics[width=60mm]{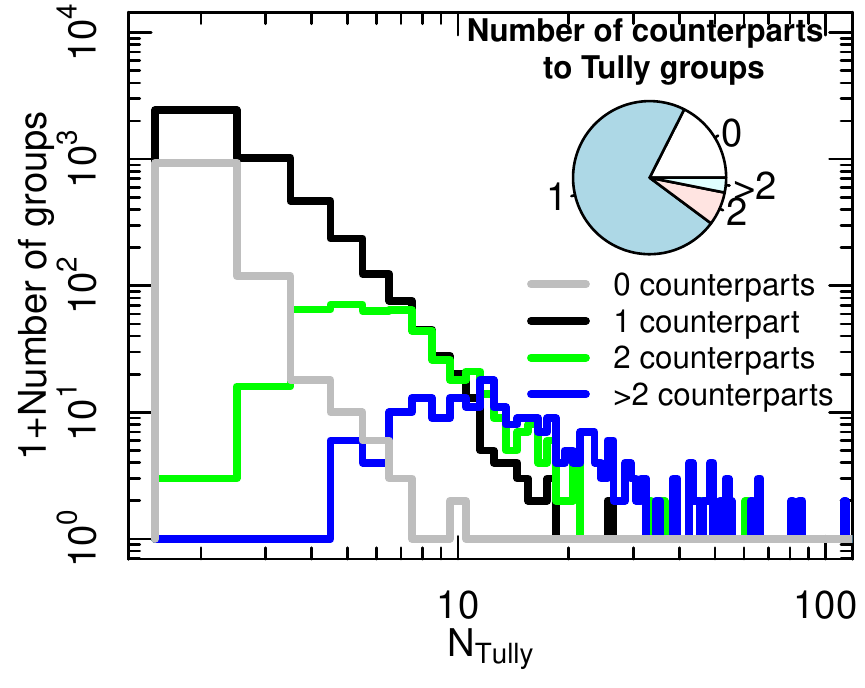} 
 \includegraphics[width=60mm]{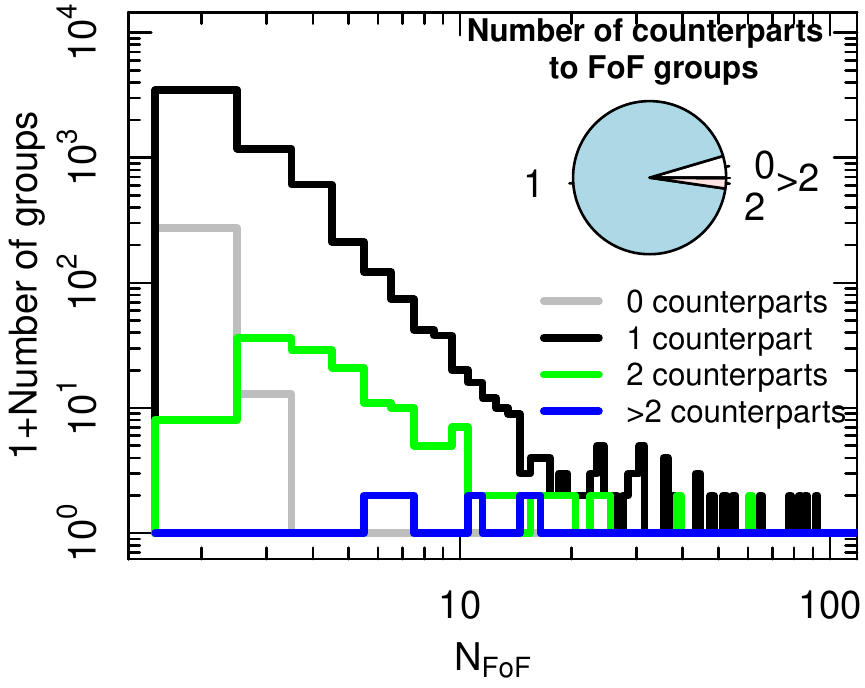}
 \caption{Distribution of matching groups as a function of group richness  in \citet{2015AJ....149..171T} and in our FoF catalogue. In the \emph{left panel}, counterparts for each Tully group are sought from the FoF catalogue, in the \emph{right panel} vice versa. Different lines correspond to groups with zero, one, two, or more matches in the comparison catalogue (see text for more details). Groups with more matches are generally richer, groups with zero matches poorer systems. The \emph{inset pie diagrams} show the fractions of groups with a given amount of counterparts. Table~\ref{tab:match} shows the numbers and fractions of groups contained in each sector.}
 \label{fig:comparison_fraction}
\end{figure*}

\begin{figure}
 \centering
 \includegraphics[width=88mm]{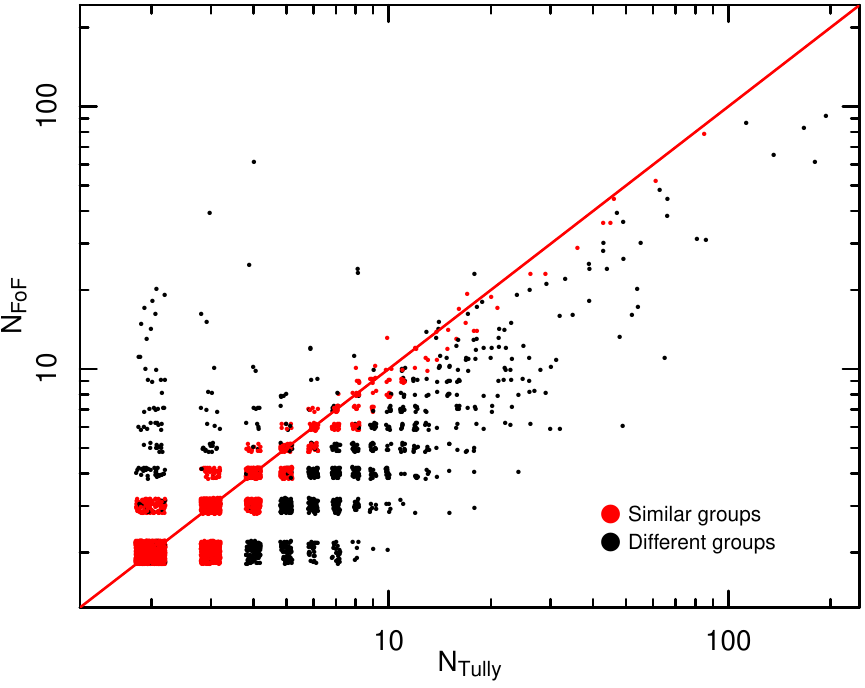}
 \caption{Group richness in \citet{2015AJ....149..171T} and in our (FoF) catalogue. For each Tully group only the best matching FoF group is shown. More than half of the groups lie on the line indicating a one-to-one match. Similar groups are shown with red colour. See text for the definitions of best match and similarity. A slight scatter is imposed on group richness for clarity.}
 \label{fig:comparison_richness}
\end{figure}

Figure~\ref{fig:dist_vs_ngal} shows the richness (number of galaxies in a group) of the detected groups as a function of distance. Farther away, galaxy systems appear to be smaller. This is a natural result for a flux-limit survey since the number density of galaxies decreases rapidly with increasing distance. The lower panel of the figure shows that nearby, the 2MRS sample provides almost as many groups as the combined dataset, while the farther end of the group sample is almost solely provided by the combined dataset. 

Figure~\ref{fig:fog} illustrates the FoG effect. In the upper panel, the observed distribution of galaxies is plotted with the observer in the centre of the data cube, at the origin of coordinates. The radial elongation of structures is clearly visible, caused by peculiar motions of galaxies in groups. To suppress these artefacts, we used velocity dispersion and projected size to spherise galaxy groups as described in Appendix~\ref{app:properties}. The lower panel in Fig.~\ref{fig:fog} shows galaxy distribution after the spherisation. Compared to the upper panel, the FoG effect is greatly reduced. An illustration of the same FoG suppression method on individual galaxy groups can be found in Fig.~6 in \citet{2012A&A...540A.106T}. This suppression cannot fully recover the true positions of galaxies in the radial direction, but reduces the average error of the distance estimates significantly. The FoG-corrected galaxy distribution is useful for several applications, for example, to construct the luminosity density field \citep[see][]{2012A&A...539A..80L} and to detect galaxy filaments \citep{2014MNRAS.438.3465T}. The latter is also one purpose of the present catalogue.

\section{Comparison with the catalogue of Tully}
\label{sec:comparison}

We compared our results with those derived by \citet{2015AJ....149..171T}. More thorough comparisons of recent grouping algorithms and their performance can be found elsewhere \citep{2014MNRAS.441.1513O, 2015MNRAS.449.1897O}.

The catalogue by \citet{2015AJ....149..171T} is the latest so far and relies on the same 2MRS dataset as was used in our analysis. On the other hand, our catalogue was constructed using a completely different approach. In contrast to our FoF algorithm, the groups in \citet{2015AJ....149..171T} were constructed using a halo-based method. This means that a dark halo is ascribed to each galaxy according to scaling relations. All galaxies lying within the boundaries of the halo are considered to belong to the same system, after which new halo parameters are calculated and the group membership is updated. This procedure is repeated iteratively until convergence.

\citet{2015AJ....149..171T} constructed his group catalogue for the full 2MRS sample, but noted that groups are only reliable within recession velocities 3000 to 10\,000~$\mathrm{km\,s^{-1}}$ (44--146~Mpc). We carried out the comparison only considering groups within this distance interval as well as using the full 2MRS sample. Qualitatively, the results are similar in both cases, while slightly better agreement is achieved within the restricted distance interval. Below we present the comparison using the full 2MRS sample.

\begin{figure*}
	\sidecaption
\includegraphics[width=12cm]{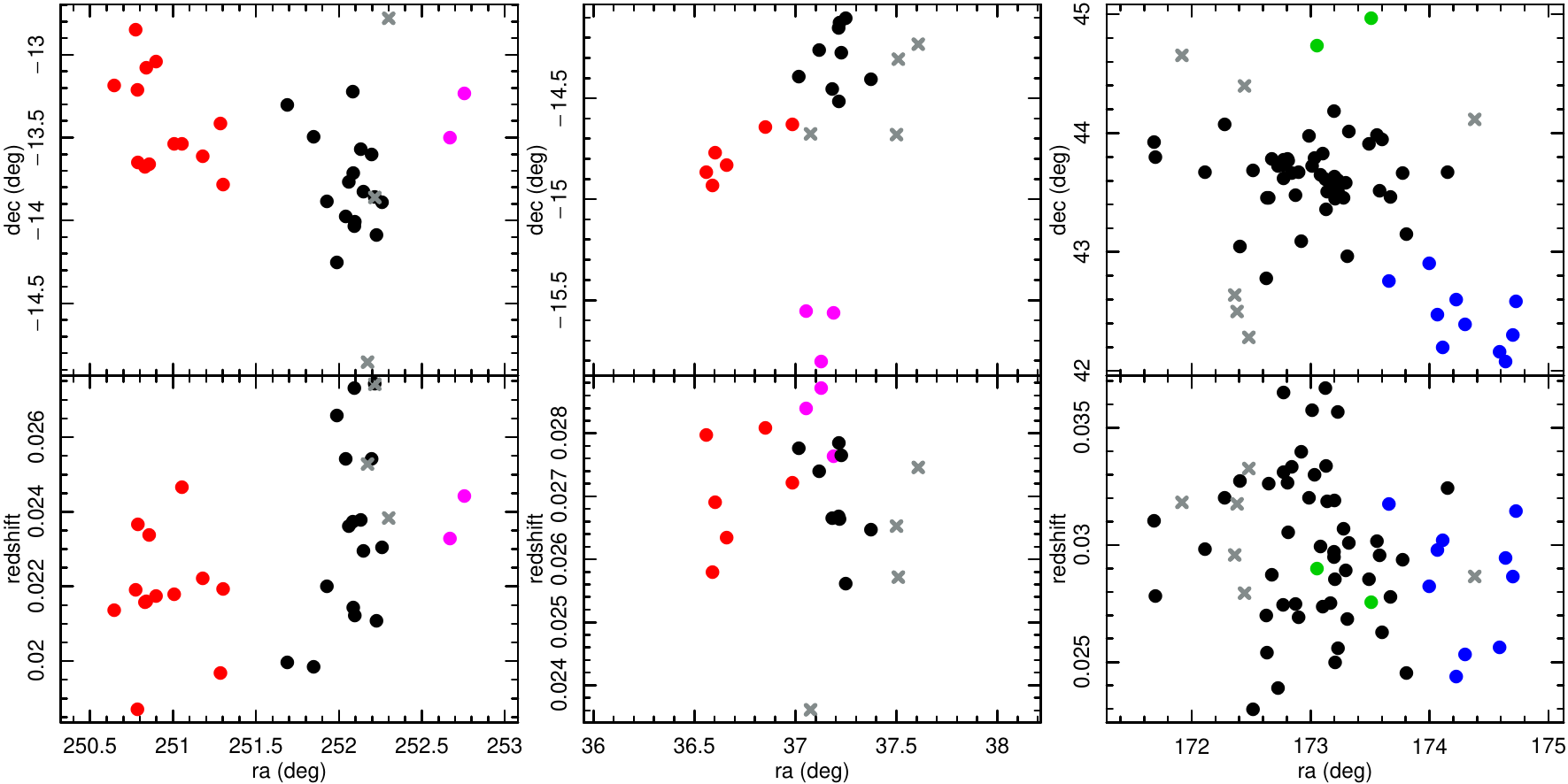}
\caption{Example of three groups in the group catalogue of \citet{2015AJ....149..171T} that dissolve into several groups in our FoF catalogue. Galaxy positions are shown in RA-Dec (\emph{upper panels}) and redshift-Dec (\emph{lower panels}) coordinates. In each panel, all galaxies belong to one Tully group. Separate colours represent separate FoF groups in our catalogue. Grey crosses designate group members in Tully catalogue that do not belong to any FoF group. They are mostly located in group outskirts and/or are separated in the redshift space. These examples show that our FoF algorithm with membership refinement suggests that a single Tully group may contain separable subcomponents and possible outliers.}
\label{fig:comparison_example}
\end{figure*}

\begin{table}
\caption{Group matching between \citet{2015AJ....149..171T} and FoF (this work) catalogues. The match is based either on the Tully or the FoF catalogue, as described in the text. The number and fraction of groups with zero, one, two or more matches in the respective comparison catalogue are shown.}
\label{tab:match}
\centering
\begin{tabular}{lrcrc}
\hline\hline
 & \multicolumn{2}{c}{\citet{2015AJ....149..171T}} & \multicolumn{2}{c}{FoF (this work)} \\
Sample & $N_\mathrm{groups}$ & Fraction & $N_\mathrm{groups}$ & Fraction \\
\hline
0 matches & 1087 & 17.5\% & 286 & \phantom{0}4.6\% \\ 
1 match & 4484 & 72.3\% & 5838 & 93.1\% \\ 
2 matches & 437 & \phantom{0}7.1\% & 137 & \phantom{0}2.2\% \\
3+ matches & 194 & \phantom{0}3.1\% & 6 & \phantom{0}0.1\% \\
\hline
Total & 6202 & 100\% & 6267 & 100\% \\
\hline
\end{tabular}
\end{table}

To conduct the comparison, we found a match between the two catalogues, referring to them as Tully and FoF catalogues below (but we recall that here the FoF groups are refined with a subsequent analysis). In the Tully catalogue we only consider groups with two or more members. The matching was made according to the membership of galaxies. For each group the best-matching group in the other catalogue is the one with the largest number of common members. This match depends on the basis catalogue. For example, we consider the case of a group in the Tully catalogue that consists of two groups in the FoF catalogue. When we match it on the basis of the Tully catalogue, we obtain one match (with the group with more members in the FoF catalogue). When we match it on the basis of the FoF catalogue, the same Tully group is matched with both FoF groups.

Figure~\ref{fig:comparison_fraction} illustrates the group matching. In the left-hand panel, the match is based on the Tully groups, in the right-hand panel on our (FoF) groups. For a majority of groups in both catalogues only a single match is found from the other catalogue. Only a few groups are split into two or more groups in the other catalogue in both cases. In slightly more cases, no match is found at all -- mostly groups with very few members. Table~\ref{tab:match} gives the number of groups in both catalogues that have zero, one, two or more matches in the comparison catalogue. If there are more than one match for a single group, the best match is considered to be the one with more members.

Figure~\ref{fig:comparison_richness} shows the richness of groups in the Tully catalogue with respect to the richness of the best-matching group in the FoF catalogue. More than half of the groups fall on the one-to-one correspondence line, which means that they are identical in both catalogues. Very similar groups are represented in red in Fig.~\ref{fig:comparison_richness}. For similarity we required that 80\% of the members in the matching groups are the same, that is, the number of galaxies in groups satisfy
$N[G_\mathrm{Tully}\cap G_\mathrm{FoF}] \geq \| 0.8 (N[G_\mathrm{Tully}\cup G_\mathrm{FoF}]) \|$, where $N[G_\mathrm{Tully/FoF}]$ indicates the number of galaxies in Tully/FoF group. Including similar groups, more than two thirds of all groups are the same in both catalogues. Taking into account that these two catalogues were constructed using completely different techniques, the correspondence is remarkable.

\begin{figure}
 \centering
 \includegraphics[width=88mm]{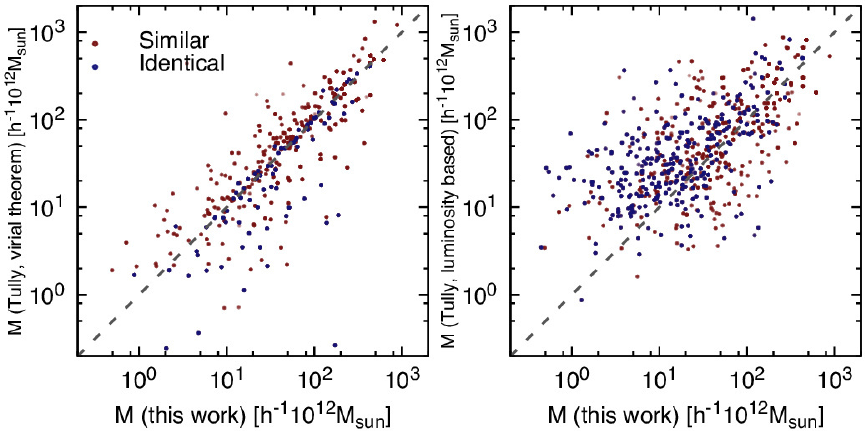}
 \caption{Comparison of group masses as estimated in this work and by \citet{2015AJ....149..171T}. We have used only the virial theorem, while \citet{2015AJ....149..171T} used the virial theorem (\emph{left panel}) and galaxy luminosities (\emph{right panel}). Here, only identical groups (blue points) and similar groups (red points) with at least four members are shown. The virial-theorem-based mass estimates agree well with each other.}
 \label{fig:mass}
\end{figure}

From the group comparison analysis, we conclude the following: most of the groups are identical in both catalogues and a majority of them are very similar. The groups that are not identical are slightly larger in Tully's catalogue than in the FoF catalogue (mostly because of membership refinement). Omitting the meaningless discussion about the true nature of galaxy groups, we can nevertheless consider this a positive result because for the FoG suppression we wish to consider each subgroup separately, keeping the modification of the observed galaxy distribution as slight as possible.

Figure~\ref{fig:comparison_example} illustrates the division of single Tully groups into multiple groups in our FoF catalogue. Three examples are shown. Subcomponents of the Tully groups are reasonably well separated in our FoF catalogue. Additionally, several galaxies in Tully groups do not belong to any group in our catalogue: galaxies in the outskirts in the plane of the sky and/or separated in the redshift space. We stress that the separation of subcomponents is not always desired; the required level of the  substructure detection largely depends on the goal of the study. To suppress the FoG (one goal of the present work), subcomponent distinction should be preferred. In general, the detection of multicomponent groups as single systems is a well-known and an often discussed topic. \citet{2015AJ....149..171T} also addressed this issue and separated the systems into two components by hand in six cases \citep[see Fig.~7 in][for an example]{2015AJ....149..171T}.

\citet{2015AJ....149..171T} estimated masses of groups using two different methods: as inferred from galaxy luminosities, and as calculated from the virial theorem. Group masses were estimated using the virial theorem in our catalogue as well, but the details of the practical application of the theorem are slightly different. See \citet{2015AJ....149..171T} and \citet{2014A&A...566A...1T} for descriptions of the two techniques.

Figure~\ref{fig:mass} compares our group mass estimates with those by \citet{2015AJ....149..171T}. Our estimates agree well with the virial theorem predictions in \citet{2015AJ....149..171T}. As expected, the luminosity-based mass estimate has a larger scatter. We conclude that regarding the differences in the virial theorem application, the masses are sufficiently concordant.

\section{Conclusions}
\label{sec:con}

We presented an improved FoF galaxy group finder with group membership refinement. In addition to the conventional FoF algorithm, we conducted a multimodality analysis to split merging groups and/or subsystems that are clearly distinguishable in the sky plane or in the redshift space. The multimodality analysis affected about half of the systems with at least ten galaxies. We refined the groups further using  virial radius and escape velocity of the groups to detect gravitationally unbound galaxies.

We applied our method on galaxies in the local Universe using two datasets: the 2MRS sample and a combined 2MRS, CF2, and 2M++ dataset. The combined dataset increases the number density of galaxies farther away. The group catalogues for both datasets are publicly available and can be accessed from \url{http://cosmodb.to.ee}.

We compared our detected groups with another recent group catalogue based on the 2MRS data by \citet{2015AJ....149..171T}. Half of the groups were found to be identical in both catalogues and two thirds are very similar. Considering that these two catalogues were constructed using completely different approaches, the agreement is remarkable. It ensures that most of the detected systems are actual galaxy groups and that both methods are meaningful. We note that the catalogues differ in details; the preference of one over the other depends on the aim of the study. Out of the non-identical groups, those in our catalogue tend to be slightly smaller and contain less substructure, thus being more favourable for studies where the FoG suppression is required.

We also compared the group masses in our catalogue and in \citet{2015AJ....149..171T}. Group masses estimated using the virial theorem agree very well in both catalogues, even though the practical application of the virial theorem has been different.

As a forthcoming application, we will use our constructed catalogue to detect galaxy filaments from the local Universe using the Bisous model \citep{2014MNRAS.438.3465T}. The data have already been used in \citet{2015MNRAS.453L.108L}, where galaxy filaments in the local Universe were shown to be well aligned with the underlying velocity field constructed using the CF2 data.

\begin{acknowledgements}
  This work was supported by institutional research funding IUT26-2, IUT40-2 and the grant ETF9428 of the Estonian Ministry of Education and Research. T.T. acknowledges financial support by the Estonian Research Council grant PUTJD5.
\end{acknowledgements}


\appendix

\section{Description of the catalogues}
\label{app:cat}

The catalogue of galaxy groups consists of two tables for both datasets (2MRS alone and combined). The first table lists galaxies that were used to generate the group catalogues, the second describes the group properties. The catalogues are available at \url{http://cosmodb.to.ee}. The catalogues will also be made available through the Strasbourg Astronomical Data Centre (CDS).

\subsection{Galaxy catalogues}

The galaxy catalogues contain the following information (column numbers are given in square brackets):
\begin{enumerate}
 \item{[1]\,\texttt{pgcid} --} identification number in PGC (principal galaxy catalogue);
 \item{[2]\,\texttt{groupid} --} group/cluster id given in the present paper;
 \item{[3]\,\texttt{ngal} --} richness (number of members) of the group/cluster the galaxy belongs to;
 \item{[4]\,\texttt{groupdist} --} comoving distance to the group/cluster centre to which the galaxy belongs, in units of Mpc, calculated as an average over all galaxies within the group/cluster;
 \item{[5]\,\texttt{zobs} --} observed redshift (without the CMB correction);
 \item{[6]\,\texttt{zcmb} --} redshift, corrected to the CMB rest frame;
 \item{[7]\,\texttt{zerr} --} error of the observed redshift;
 \item{[8]\,\texttt{dist} --} comoving distance in units of Mpc (calculated directly from the CMB-corrected redshift);
 \item{[9]\,\texttt{dist\_cor} --} comoving distance of the galaxy after suppressing the finger-of-god effect (see Appendix~\ref{app:properties});
 \item{[10--11]\,\texttt{raj2000, dej2000} --} right ascension and declination (deg);
 \item{[12--13]\,\texttt{glon, glat} --} Galactic longitude and latitude (deg);
 \item{[14--15]\,\texttt{sglon, sglat} --} supergalactic longitude and latitude (deg);
 \item{[16--18]\,\texttt{xyz\_sg} --} supergalactic cartesian coordinates in units of Mpc based on dist\_cor (fingers of god are suppressed);
 \item{[19]\,\texttt{mag\_ks} --} Galactic-extinction-corrected $K_s$ magnitude as given in source catalogue;
 \item{[20]\,\texttt{source} --} source of the galaxy: 1 for 2MRS, 2 for CF2, 3 for 2M++.
\end{enumerate}

\subsection{Description of group catalogues}

The catalogues of groups/clusters contain the following information (column numbers are given in square brackets):
\begin{enumerate}
 \item{[1]\,\texttt{groupid} --} group/cluster id;
 \item{[2]\,\texttt{ngal} --} richness (number of members) of the group;
 \item{[3--4]\,\texttt{raj2000, dej2000} --} right ascension and declination of the group centre (deg);
 \item{[5--6]\,\texttt{glon, glat} --} Galactic longitude and latitude of the group centre (deg);
 \item{[7--8]\,\texttt{sglon, sglat} --} supergalactic longitude and latitude of the group centre (deg);
 \item{[9]\,\texttt{zcmb} --} CMB-corrected redshift of the group, calculated as an average over all group/cluster members;
 \item{[10]\,\texttt{groupdist} --} comoving distance to the group centre (Mpc);
 \item{[11]\,\texttt{sigma\_v} --} rms deviation of the radial velocities ($\sigma_v$ in physical coordinates, in \mbox{km\,s$^{-1}$});
 \item{[12]\,\texttt{sigma\_sky} --} rms deviation of the projected distances in the sky from the group centre ($\sigma_\mathrm{sky}$ in physical coordinates, in Mpc), $\sigma_\mathrm{sky}$ defines the extent of the group in the sky;
	\item{[13]\,\texttt{r\_max} --} distance (in Mpc) from group centre to the farthest group member in the plane of the sky;
 \item{[14]\,\texttt{mass\_200} --} estimated mass of the group assuming the NFW density profile (in units of $10^{12}M_\odot$);
	\item{[15]\,\texttt{r\_200} --} radius (in kpc) of the sphere in which the mean density of the group is 200 times higher than the average of the Universe;
	\item{[16]\,\texttt{mag\_group} --} observed magnitude of the group, i.e. the sum of the luminosities of the galaxies in the group.
\end{enumerate}

\section{Basic properties of galaxy groups}
\label{app:properties}

For every galaxy group we calculate several basic properties. The main properties are the velocity dispersion and the size in the plane of the sky, used for estimating the virial mass and radius of the groups and for the suppression of the FoG redshift distortions. Details about these calculations are given in \citet{2014A&A...566A...1T}. For convenience a condensed description is provided below.

The group velocity dispersion $\sigma_v^2$ is calculated with the formula
\begin{equation}
	\sigma_v^2 = \frac{1}{(1+z_\mathrm{m})(n-1)}\sum\limits_{i=1}^{n}(v_i-v_\mathrm{m})^2,
	\label{eq:sigvel}
\end{equation}
where $z_\mathrm{m}$ and $v_\mathrm{m}$ are the mean redshift and velocity of the group; $v_i$ are the velocities for individual group members. Summation is over all galaxies with a measured velocity within the group.

The group extent in the sky plane is defined as
\begin{equation}
	\sigma_\mathrm{sky}^2 = \frac{1}{2n(1+z_\mathrm{m})^2}\sum\limits_{i=1}^{n}(r_i)^2,
	\label{eq:sigsky}
\end{equation}
where $r_i$ are the projected distances (in comoving coordinates) from the group centre in the plane of the sky.

Both quantities, velocity dispersion $\sigma_v^2$ and group extent $\sigma_\mathrm{sky}^2$, are defined in physical units. This is an obvious choice since we use these quantities to calculate the physical properties of groups, the virial mass and radius.

Group masses are estimated using the virial theorem from which we can derive the equation
\begin{equation}
	M_\mathrm{vir} = 2.325\times 10^{12}\frac{R_g}{\mathrm{Mpc}}\left(\frac{\sigma_v}{100\,\mathrm{km\,s}^{-1}}\right)^2M_\sun,
\end{equation}
where $R_g$ is the gravitational radius, which for a fixed mass density profile only depends on the group extent in the sky $\sigma_\mathrm{sky}^2$. To estimate group masses we assume an NFW profile \citep{1997ApJ...490..493N} using mass-concentration relation as derived in \citet{2008MNRAS.391.1940M}. As a result, the NFW profile only depends on the mass. See \citet{2014A&A...566A...1T} for details about gravitational radius and mass calculations. Under the assumption of an NFW profile, the group virial radius is uniquely defined with the virial mass. The virial radius is defined as the radius in which the mean density is 200 times higher than the mean density in the Universe.

To suppress the FoG redshift distortions we use the rms sizes of groups in the sky ($\sigma_\mathrm{sky}$) and their rms radial velocities ($\sigma_v$). Both are given in physical units as defined above. To suppress redshift distortions, we calculate new radial distances for galaxies using the formula
\begin{equation}
	d_\mathrm{gal} = d_\mathrm{group} + \left(d^*_\mathrm{gal}-d_\mathrm{group}\right)\frac{\sigma_\mathrm{sky}}{\sigma_v/H_0},
\end{equation}
where $d^*_\mathrm{gal}$ is the initial distance (calculated directly from galaxy redshift) to the galaxy, $d_\mathrm{group}$ is the distance to the group centre, and $H_0$ is the Hubble constant. For galaxy pairs, we demand that their size along the line of sight does not exceed the linking length $d_\mathrm{LL}(z)$ used to define the system
\begin{eqnarray}
 d_{\mbox{gal}}&=&d_{\mbox{group}}+\left(d^{\star}_{\mbox{gal}}-d_{\mbox{group}}\right)\frac{d_{LL}(z)}{|v_1-v_2|/H_0},\\
 &&\mbox{if}\;|v_1-v_2|/H_0 > d_{LL}(z).\nonumber
\end{eqnarray}
Here $z$ is the mean redshift of a galaxy pair.

The suppression of FoG redshift distortions as defined above was initially used to calculate the luminosity density field using SDSS data \citep[see][]{2012A&A...539A..80L} and has later been successfully used to prepare the data for galaxy filament detection \citep{2014MNRAS.438.3465T}.

\end{document}